\documentclass{ws-dmaa}

\usepackage{color,array}
\usepackage{graphicx}
\usepackage{amsmath,amsfonts}
\usepackage[noend]{algpseudocode}
\usepackage{algorithmicx,algorithm}
\usepackage{array}
\usepackage[caption=false,font=normalsize,labelfont=sf,textfont=sf]{subfig}
\usepackage{textcomp}
\usepackage{stfloats}
\usepackage{url}
\usepackage{verbatim}
\usepackage{graphicx}
\usepackage{longtable}
\usepackage{tabularx}

\usepackage{booktabs, makecell, tabularx}
\usepackage{stfloats}
\usepackage{siunitx}
\usepackage{multirow}

\usepackage[misc,geometry]{ifsym}

\usepackage[symbol]{footmisc}

\newcolumntype{M}[1]{>{\centering\arraybackslash}m{#1}}

\begin{document}

\markboth{T. Chen et al.}
{Graph Representation Learning for Popularity Prediction Problem: A Survey}

%
\catchline{}{}{}{}{}
%

\title{Graph Representation Learning for Popularity Prediction Problem: A Survey
}






\author{Tiantian Chen$^{*,\P,\text{\Letter}}$\footnotetext{\Letter\ Corresponding author}, Jianxiong Guo$^{\dag, \S}$ and Weili Wu$^{*, \ddag}$}

\address{$^{*}$Department of Computer Science, University of Texas at Dallas\\
800 W Campbell Rd, Richardson, TX 75080, USA.\\
$^{\dag}$Advanced Institute of Natural Sciences, Beijing Normal University\\ 
Zhuhai 519087, China\\
Guangdong Key Lab of AI and Multi-Modal Data Processing, BNU-HKBU United International College, Zhuhai 519087, China\\
$^{\P}$tiantian.chen@utdallas.edu\\
$^{\S}$jianxiongguo@bnu.edu.cn\\
$^{\ddag}$weiliwu@utdallas.edu\\
}

\maketitle

\begin{history}
\received{Day Month Year}
\revised{Day Month Year}
\accepted{Day Month Year}
\published{Day Month Year}
\end{history}

\begin{abstract}
The online social platforms, like Twitter, Facebook, LinkedIn and WeChat, have grown really fast in last decade and have been one of the most effective platforms for people to communicate and share information with each other. Due to the “word-of-mouth” effects, information usually can spread rapidly on these social media platforms. Therefore, it is important to study the mechanisms driving the information diffusion and quantify the consequence of information spread. A lot of efforts have been focused on this problem to help us better understand and achieve higher performance in viral marketing and advertising.  On the other hand, the development of neural networks has blossomed in the last few years, leading to a large number of graph representation learning (GRL) models. Compared with traditional models, GRL methods are often shown to be more effective. In this paper, we present a comprehensive review for recent works leveraging GRL methods for popularity prediction problem, and categorize related literatures into two big classes, according to their mainly used model and techniques: embedding-based methods and deep learning methods. Deep learning method is further classif\mbox{i}ed into convolutional neural networks, graph convolutional networks, graph attention networks, graph neural networks, recurrent neural networks, and reinforcement learning. We compare the performance of these different models and discuss their strengths and limitations. Finally, we outline the challenges and future chances for popularity prediction problem.   
\end{abstract}



\section{Introduction}

Online social platforms, such as Twitter, Facebook and WeChat, have become one of the most effective ways for people to obtain information and to communicate with each other. Due to the "word-of-mouth" effects, information can spread fast on social networks. Many companies have utilized this to promote their products by distributing some free samples or coupons to some influential users on the networks. There are lots of works focused on studying diffusion on networks, including diffusion of news, adoptions of new products, etc. One of the extensively studied topics is influence maximization (IM) problem \cite{KempeKT03, BorodinFO10, ChenWY09, ChenWW10, ChenYZ10}, seeking for a small number of nodes as seeds to spread information on social networks to maximize the total number of activated nodes, under some diffusion model. For example, two commonly used diffusion models in existing works are: Independent Cascade (IC) model and Linear Threshold (LT) model \cite{KempeKT03}. IC model focuses more on the peer-to-peer communication, while in LT model node incorporates the influence from its in-neighbors.      

It has been proved that IM problem is NP-hard and computing the influence spread $\sigma(S)$ of a node set $S$ under both the IC and LT model is $\#$P-hard \cite{ChenWW10, ChenYZ10}. Kempe \textit{et al.} \cite{KempeKT03} proposed the greedy algorithm for the IM problem and utilized the Monte Carlo method to estimate the influence spread of a seed set, which can achieve $(1-1/e-\epsilon)$-approximation ratio. However, Monte Carlo is too time-consuming and incurs signif\mbox{i}cant computation overheads. Many works have been focused on solving this problem \cite{KempeKT05,LeskovecKGFVG07, ChenWY09, ChenYZ10, GoyalLL11,abs-1111-4795,GalhotraAR16}. But most of these approaches either trade performance guarantees for practical eff\mbox{i}ciency, or vice versa. There are some exceptions, like \cite{TangXS14, TangSX15, Chen18, NguyenTD16, HuangWBXL17}, based on a novel Reverse Influence Sampling (RIS) technique, which can not only obtain a $(1-1/e)$-approximation solution with high probability but be eff\mbox{i}cient even for large-scale datasets. Later, a large number of variants of IM problem are developed \cite{GuoW20, ZhuGZW19, ZhuLYWB17,ZhuGW19}, such as prof\mbox{i}t maximization \cite{ChenLLFYW20, LiuLWFDW20,abs-1802-06946, ZhuGWG19}, activity maximization \cite{YangYWMD19, GuoCW20a}, rumor blocking \cite{TongWGLLLD20, YanLWDW20, GuoLW20,TongDW18, YanLWLW19, ZhuGW21, ZhangYWL20, TongLWD16,FanLWTMB13, TongWD182}, community detection \cite{ZhangGYW21,WagensellerWW18, YuanWV18, TongCWLD16}, and adaptive influence maximization problem \cite{GuoW21, ton/TongWTD17,abs-2004-00893}. Objectives of these problems are still hard to estimate and the proposed algorithms are basically approximation algorithms based on RIS technique or heuristics without any approximation guarantee.

On the other hand, for these IM and its variants problem, we need assume the information spreads on the networks according to some diffusion models. Even though IC and LT model are frequently used in previous works, information diffusion on real networks are more complex than either the IC model or the LT model. Based on IC and LT model, many variant diffusion models have been proposed to simulate more practical and complicated scenarios \cite{LiuCXZ12, ChenLZ12, GuoCW21, abs-2006-03222, ShanCLSZ19, LeeKY12, ZhangYGY19}, which still have some limitations. Therefore, understanding how the information is diffused on the networks and which factor can make the information spread successfully on the networks, and predicting the size of the popularity are challenging but important in many practical applications, like advertising \cite{GuoCW20, GuoW202}, viral marketing \cite{GuoW19, tcss/GuoW20}, and recommendation \cite{kdd/WuGGWC19}. 

The diffusion process and trajectories of information propagation as well as the participants in the diffusion of information are called the information cascade. Many works have been focused on learning information diffusion problems in social networks \cite{yang2021full, FengCKLLC18, wang2020learning, Chen0ZTZZ19, QiuTMDW018, LiMGM17, abs-1812-08933, Yuan18arxiv, CaoSGWC20, XiaLWL21, ChenZ0TZZ19}. One of the most interesting and important problems related to information cascade is the popularity prediction. Given some observed cascade data (usually part of the diffusion result with/without timestamp), the popularity prediction problem aims to predict the f\mbox{i}nal size of the cascade based on observations, or more specif\mbox{i}cally, the activation probability that a user is activated at the end of the diffusion process.  

In this paper, we provide a comprehensive review of existing works related to popularity prediction problems. This is a challenging task because there are a large number of related works and there is no uniform standard to classify these works. According to the prediction level, popularity prediction problem usually can be classif\mbox{i}ed into three types:  micro-level, meso-level and macro-level prediction. Microscopic popularity prediction focuses more on who will be the next activated user and maybe also the activation time of it, while macroscopic popularity prediction cares more about how many users will be activated within a specif\mbox{i}c time interval and the total number of active users when the information propagation process is stopped. Meso-level prediction focuses on the behaviors of clusters or communities. From the methodology view, there are many choices of algorithmic methods for modeling and predicting information cascades. Traditional machine learning methods, like linear/logistic regression and Support Vector Machine (SVM), focus on extracting various temporal and structural features associated with information items by feature engineering. However, these methods have some limitations since these features cannot be adapted during the learning process and obtaining them can be time-consuming and expensive \cite{HamiltonYL17}. With  recent advancements of deep neural networks, especially the techniques for graph representation learning (GRL), like DeepWalk \cite{PerozziAS14}, node2vec \cite{GroverL16}, graph convolutional networks \cite{KipfW16}, recurrent neural networks and its variants \cite{HochreiterS97, ChungGCB14}, many deep learning (DL) models for popularity prediction problem have emerged \cite{Yuan18arxiv, CaoSGWC20, XiaLWL21, ChenZ0TZZ19}.

Several recent works \cite{GaoCLYCT19, GuilleHFZ13, ZhouXTZ21} have been focused on reviewing papers related to  popularity prediction problem. Guille \textit{et al.} \cite{GuilleHFZ13}  focused on various feature engineering methods and classic machine learning approaches for modeling and predicting the information popularity. Gao \textit{et al.} \cite{GaoCLYCT19} concentrated on microblog information propagation, while emphasizing different aspects of information cascade. Zhou \textit{et al.} \cite{ZhouXTZ21} focused more on the general perspective of the popularity prediction problem, and literatures using GRL methods especially recent works are not comprehensive and complete in their paper. However, this paper is more focused on papers using GRL method to solve the popularity prediction problem, and talks about more details about those methods. 

Contributions of this paper are as follows:
\begin{itemize}
\item We provide a comprehensive review for recent literatures using graph representation learning methods to solve the popularity prediction problem.
\item To better compare different models, we propose a new taxonomy for classifying graph representation learning methods, which basically include two main types: embedding-based methods and deep learning methods.
\item We compare the strengths and weaknesses of different models, and provide challenges and chances for future works.
\end{itemize}

\begin{figure*}[htbp]
  \centering
  \includegraphics[width=0.9\linewidth]{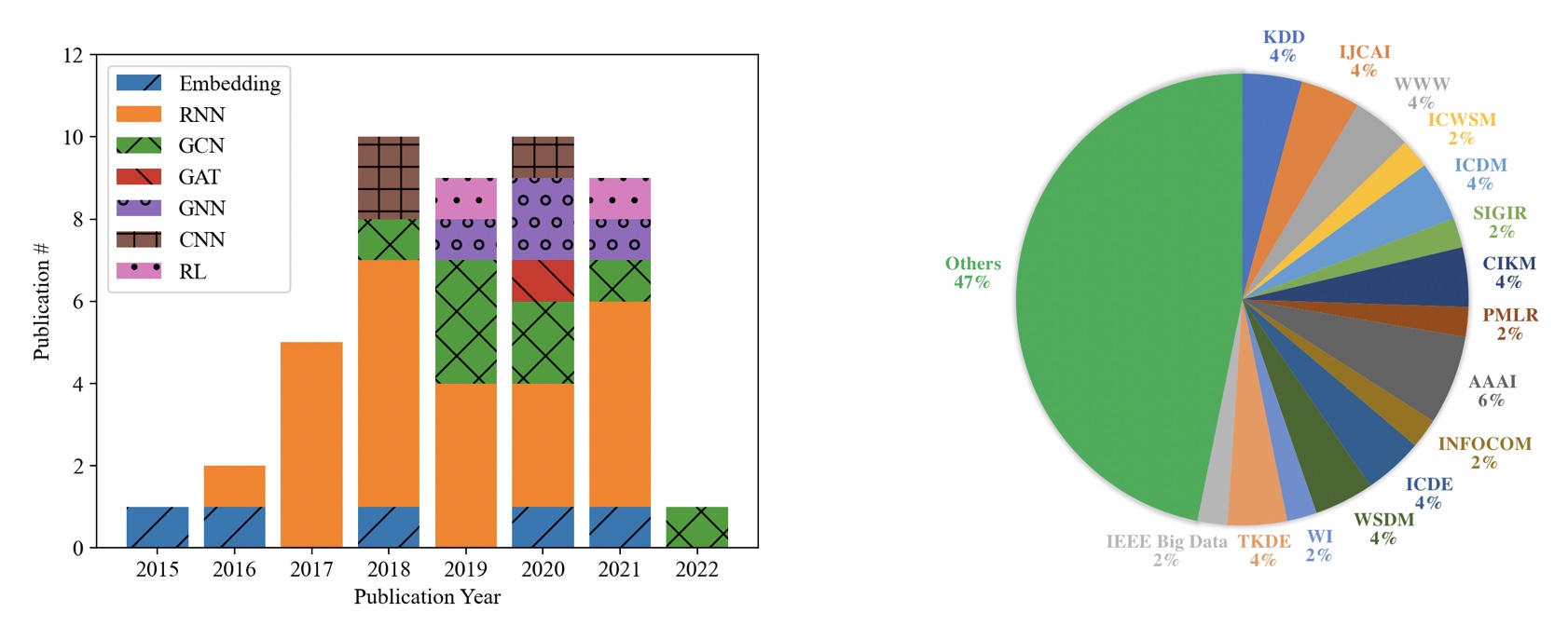}
\caption{Left: Number of publications over the last decade. Right: Source distribution of publications.} 
  \label{summary}
\end{figure*}

Fig. \ref{summary} shows a statistics summary of papers related to popularity prediction problem reviewed in this survey. It can be seen that popularity prediction has attracted more attention since 2018. Among those publications, more than half of the reviewed papers are from high-ranking conferences and journals, such as WWW, IJCAI, KDD, AAAI, CIKM, ICDE, TKDE, and SIGIR.

\noindent
\textbf{Organizations}. In Section \ref{Preliminaries}, we f\mbox{i}rst introduce some basic def\mbox{i}nitions that will be used frequently in following parts. 
Section \ref{sumrew} is dedicated to summarizing existing literatures using graph representation learning methods for popularity prediction.
Section \ref{Embedding} and Section \ref{DeepL} review the embedding-based methods and deep learning methods for popularity prediction problem, respectively. In Section \ref{Discussion}, we discuss advantages and disadvantages of these models, and outline challenges and future opportunities for popularity prediction problem. Section \ref{conclusion}  concludes this paper.

\begin{table}[hptb]\centering
  \caption{Def\mbox{i}nitions of abbreviations.}
  \label{abbreviation}
  \begin{tabular}{c|c}
    \hline
    Abbreviation & Def\mbox{i}nition\\
    \hline
    GRL & Graph representation learning\\
    DL & Deep learning\\
    CNN & Convolutional neural network\\
    GCN & Graph convolutional network\\
    GAT & Graph attention network\\
    GNN & Graph neural network\\
    RNN & Recurrent neural network\\
    LSTM & Long short-term memory\\
    GRU & Gated recurrent units\\
    RL & Reinforcement learning\\
    \hline
\end{tabular}
\end{table}

\section{Preliminaries}\label{Preliminaries}
The diffusion process of information on social networks is complicated and it is important to understand how information spreads on networks. In this section, we will introduce some basic def\mbox{i}nitions related to popularity prediction. Table \ref{abbreviation} lists some abbreviations used in this survey.

\subsection{Social network}
With the fast development of Internet and software, many kinds of contents, like text, image and video, can spread through websites, phone applications, etc. For example, one user in Twitter posts a tweet including texts and photos, and then all of his followers can see the tweet. Later, some of his followers may retweet this tweet. We refer to all of this kind of contents/items that can spread among users as information. In that case, information itself may have many features. But in this survey, we do not focus on those papers considering contents features of the information, but focus more on papers that use different models to capture structural and temporal features from cascades.

Users and relationships between them play important roles in information propagation, which construct a social network. There are many types of social network, such as friendship network and citation network. A social network is usually denoted by a directed graph $G=(V, E)$, where $V$ is the set of nodes (users) and $E$ represents the set of relationships between users. For edge $(u, v)\in E$, $u$ is called the in-neighbor of $v$ and $v$ is the out-neighbor of $u$. 

\subsection{Information cascade}
When information spreads on social network, it will start from one or a few users, and propagate to their out-neighbors and out-neighbors' out-neighbors, which will generate the information cascade. Once a user is influenced by the information, its state will change from inactive to active, and we say that it is activated. Its state will remain active and does not change in the following propagation process.

\begin{definition}[Cascade]
A cascade $\mathcal{C}_{T}=\{u_{1}, \ldots, u_{l}|u_{i}\in V, 1\leq i\leq l\}$ is a sequence of nodes ordered by the timestamp that they are activated within the observation time window $T$.  
\end{definition}

\begin{definition}[Timestamped Cascade]
A timestamped cascade $\mathcal{R_{T}}=\{(u_{1}, t_{u_{1}}), \ldots, (u_{l}, t_{u_{l}})\}$ is a sequence of tuples $(u_{i}, t_{u_{i}})$, where $t_{u_{i}}$ is the time that node $u$ is observed activated within the observation time window $T$. 
\end{definition}

\begin{definition}[Cascade Diffusion Trajectory]
A cascade diffusion trajectory $\mathcal{M_{T}}=\{(u_{1}, v_{1}, t_{1})$, $\ldots, (u_{l}, v_{l}, t_{l})\}$ is a sequence of tuples $(u_{i}, v_{i}, t_{i})$, representing that $v_{i}$ is activated by $u_{i}$  via edge $(u_{i}, v_{i})$ at time $t_{i}$. A cascade diffusion trajectory is the concrete diffusion path of the information.
\end{definition}

\begin{definition}[Cascade Graph]
A cascade graph is a snapshot of the social network at some observation time $T$ with respect to the information diffusion and is denoted as $\mathcal{G}=(\mathcal{V}, \mathcal{E}, T)$, where $\mathcal{V}$ contains all nodes that have been activated by the information by time $T$ and $\mathcal{E}\subseteq (\mathcal{V} \times \mathcal{V})$ represents the exact retweet or citation diffusion of the information. Given a diffusion trajectory of a cascade, we are able to construct its corresponding cascade graph.
\end{definition}

\subsection{Popularity prediction}
For a piece of information, we refer to \textit{popularity} as the total number of nodes activated by it when its propagation stops. Take the tweets as an example. The popularity of a tweet is def\mbox{i}ned as the total number of its retweets. For the given information $I$, denote by $P_{I}^{t}$ the number of nodes activated by it by time $t$. When no time is specif\mbox{i}ed, it refers to the f\mbox{i}nal popularity $P_{I}^{\infty}$ and we denote it as $P_{I}$. Given some cascade observations, the goal of the popularity prediction problem is to predict the f\mbox{i}nal influence or attention of this information.

\begin{figure*}[htbp]
  \centering
  \includegraphics[width=0.95\linewidth]{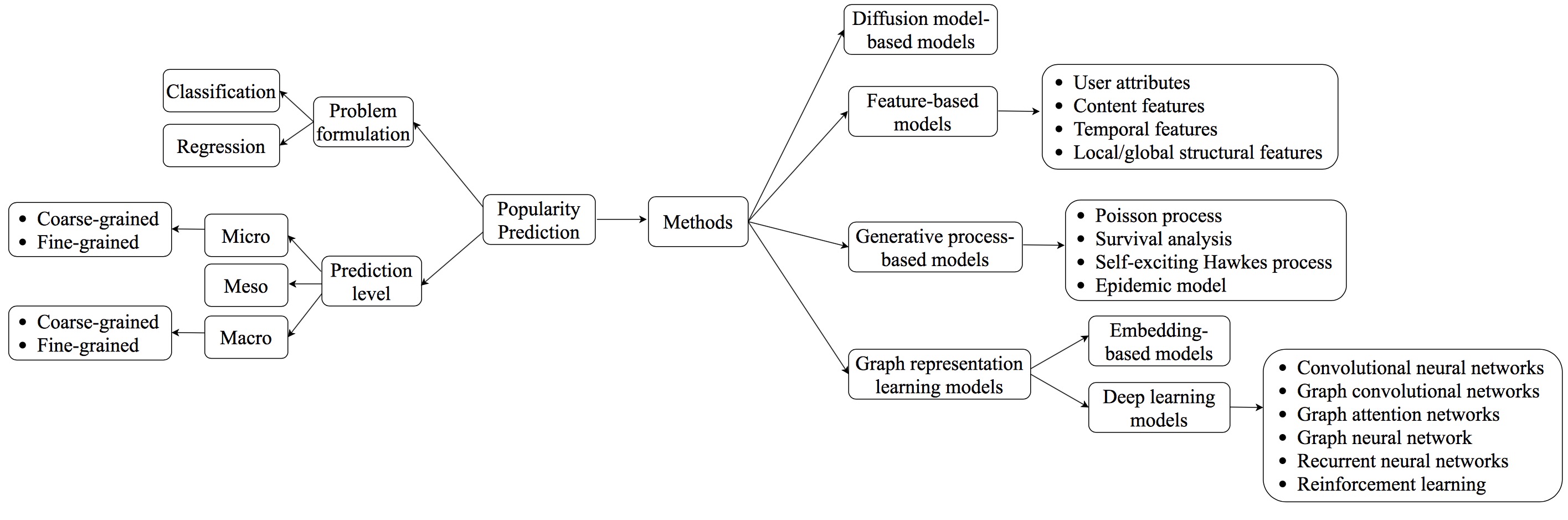}
  \caption{Category of popularity prediction models.} 
  \label{category}
\end{figure*}

In Fig. \ref{category}, we show the category of models for popularity prediction problem.  According to the problem formulation, popularity prediction problem usually can be classif\mbox{i}ed into two types: classif\mbox{i}cation and regression. For the regression problem, we need to predict the exact value of the f\mbox{i}nal popularity $P_{I}$ or the popularity of the information at a specif\mbox{i}c time $t$. However, for many practical scenarios, the exact count of the f\mbox{i}nal popularity is not required and we just need predict if the information will be popular. For example, when a company wants to promote a new product, it is crucial to predict if the product will be popular or not before the seeding strategy is implemented to save cost and obtain more prof\mbox{i}ts. Usually, given a predef\mbox{i}ned threshold $\theta$, the classif\mbox{i}cation task aims to predict whether f\mbox{i}nal popularity $P_{I}$ of the information will be more than $\theta$ or not.  

The regression task can be further categorized into three levels based on the granularity: micro-level, meso-level and macro-level. Micro-level prediction aims at the statuses of individuals, such as the activation probability of a specif\mbox{i}c user, while macro-level prediction focuses on the general outcome of the whole cascade. Meso-level prediction is focused on the behaviors of clusters or communities. Yang \textit{et al.} \cite{yang2021full} further classif\mbox{i}ed macro-level and micro-level prediction into two sub-versions: coarse-grained and f\mbox{i}ne-grained. Specif\mbox{i}cally, f\mbox{i}ne-grained macroscopic prediction focuses on how many nodes will be activated within a specif\mbox{i}c time interval, while coarse-grained version aims to estimate the f\mbox{i}nal popularity of the information, i.e., the total number of activated users by the end of the information diffusion process. For the micro-level prediction, the f\mbox{i}ne-grained version aims to predict who is the next activated node and is activated at what time, while the coarse-grained version ignores the exact activation time and only predicts the next node that will be activated.

\subsection{Learning methods}\label{learningMethods}

Methodologically, existing methods of popularity prediction problem can be classif\mbox{i}ed into diffusion model-based methods, feature-based methods, generative process-based methods and GRL methods. For GRL methods, we further classif\mbox{i} them into two types: embedding-based methods, and deep learning (DL) methods. In this paper, we will mainly focus on reviewing the GRL methods for the popularity prediction problem. 

To estimate the popularity of the information, early works usually assumed an underlying diffusion model, such as IC model that assumes there is an independent propagation probability between each pair of users. Instead of assuming diffusion probabilities, diffusion model-based methods \cite{Gomez-RodriguezLK12, WangHYL14} try to infer propagation probabilities between users, which can be represented as a network of users (also called "network inference"). However, strong hypotheses bases are needed in these methods.

Feature-based approaches concentrate on extracting various handcrafted features from raw data, which include user attributes, content features, temporal features and local/global structural features, and then formulate the popularity prediction problem as a classif\mbox{i}cation or regression task. Afterward, machine learning techniques, such as linear regression, SVM and decision tree, are adopted for the regression or classif\mbox{i}cation. Many works \cite{JendersKN13, KongMFYZ14, WuR14, Yang0MY14} have been focused on feature-based methods. \cite{ZhouXTZ21} and \cite{GaoCLYCT19} have more detailed classif\mbox{i}cation about using feature-based methods to solve popularity prediction. The predictive performance of feature-based methods is highly dependent on the quality of engineered features. Such features are often carefully designed and not easy to generalize to a different platform or domain. Feature-based methods are limited since these handcrafted features cannot be adapted during the learning process and obtaining them might be expensive and time-consuming.

In contrast, generative process-based approaches aim to model users' activation process as a point process and explore the underlying mechanisms driving the propagation dynamics of information cascade.  The commonly used models for diffusion process include Poisson process \cite{ShenWSB14}, survival analysis \cite{LermanH10, LeskovecAH07}, self-exciting Hawkes process and Epidemic model \cite{ZhaoEHRL15, mcpherson2001birds}. The point process is supposed to f\mbox{i}t the real growth curve of cascade by a carefully designed conditional intensity function. Even though these works make the cascade prediction more interpretable, they ignore the implicit information in the propagation dynamics. Besides, their predictive power is also limited because they are not proposed and optimized for popularity prediction.

The propose of GRL solves the feature extraction problem in feature-based methods, which can automatically learn the features and alleviate the need to do feature engineering every single time. The goal of GRL is to learn a mapping from nodes, edges or the entire graph, to low-dimensional vectors in $\mathbb{R}^{d}$, where geometric relationships in latent space $\mathbb{R}^{d}$ correspond to relationships in the original graph. After optimizing the embedding space, the learned embeddings can be employed as feature input for downstream machine learning tasks. The main difference between previous works and GRL is the way they deal with representing graph structure. In previous works, representing graph structure is treated as a pre-processing step, which captures the structural information by extracting handcrafted features. However, GRL methods consider it as a machine learning task itself and apply data-driven approaches to learn embeddings that encode information of graph structure \cite{HamiltonYL17}.

However, simple embedding-based approaches based on DeepWalk \cite{PerozziAS14} and node2vec \cite{GroverL16} fail to capture the time ordering of users being activated for predicting the next activated user, and have been demonstrated to be suboptimal options in experiments of recent deep-learning based methods. Recently, many DL methods, especially RNN-based models, have shown their effectiveness in popularity prediction. We will introduce these methods in the next section.

\section{Summary of GRL Models}\label{sumrew}
 The GRL models present a new way of approximating traditional graph convolutionals and kernels, allowing for faster embedding extraction. Mixing these methods with the supervised or semi-supervised methods can give the state-of-the-art models with respect to scalability, speed and quality on downstream tasks. Besides, heavy feature engineering is not needed in representation learning models which can capture non-linear representations of users and popularity spread dynamics.

\setlength{\tabcolsep}{0.0pt}

{\renewcommand{\arraystretch}{1.37}%
    \begin{table}[hptb]
    \centering
    \caption{Summary of different methods}
\label{tab:freq}
{\begin{tabular}{c||M{1.8em}|M{1.8em}|M{1.8em}|M{1.8em}|M{1.8em}|M{1.8em}|M{1.8em}||M{2.1em}||M{1.8em}|M{1.8em}|M{1.8em}||c}
    \toprule
    \multirow{2}{*}{Method} & \multicolumn{7}{c||}{Inputs} &\multirow{2}{*}{C/R} &\multicolumn{3}{c||}{Level}  & \multirow{2}{*}{GRL Model}\\
    \cline{2-8}
    \cline{10-12}
    & C & NC & T & CT & SN& CS & CG & & Mi &Ma &Me & \\
    \hline
    \hline
    Embedded IC \cite{BourigaultLG16}&$\surd$&&$\surd$&&&&&R&C&&&Embedding\\
    \hline
    LIS \cite{WangSLC15} &$\surd$&&&&&&&R &C& && Embedding\\
    \hline
    Inf2vec \cite{FengCKLLC18}&$\surd$&&$\surd$&&$\surd$& &&R & C&& & Embedding\\
    \hline
     IMINFECTOR \cite{panagopoulos2020multi}&$\surd$&&&&&$\surd$&&R&C&C&& Embedding\\
    \hline
    RNe2Vec \cite{computing/ShangHZPLLX21}&&&&$\surd$&&&&C&&&&Embedding\\
    \hline
    \hline
    Cas2vec \cite{KefatoSBSMG18}&&&$\surd$&&&&&C&&&&CNN+PM\\
    \hline
    NDM \cite{abs-1812-08933}&$\surd$&&&&&&&R&C&&&CNN+AM\\
    \hline
    HDD \cite{MolaeiZV20}&&&&$\surd$&&&&R&C&&&CNN+LSTM\\ 
    \hline
    \hline
    DeepInf \cite{QiuTMDW018}&&$\surd$&&&$\surd$&&&R&C&&&GCN/GAT\\
    \hline
    CasCN \cite{Chen0ZTZZ19}&&&&$\surd$&&&&R&&C&&GCN+LSTM\\
    \hline
    DyHGCN \cite{YuanLZLZH20}&$\surd$&&$\surd$&&$\surd$&&&R&F&&&GCN\\
    \hline
    CasSeqGCN \cite{abs-2110-06836}&&&$\surd$&&$\surd$&&$\surd$&R&&F&&GCN+LSTM\\
    \hline
    HPPNP \cite{LeungCMD019}&&$\surd$&&&$\surd$&&&R&C&&&GCN/GAT\\
    \hline
    \cite{ChungHC19}&&&&&&&$\surd$&R&C&&&GCN+AM\\
    \hline
     Multi-Influor \cite{wang2020learning}&$\surd$&&$\surd$&&$\surd$&&&R&F&&&GCN\\
    \hline
    \multirow{2}{*}{CasGCN\cite{corr/abs-2009-05152}}&&&&&&&\multirow{2}{*}{$\surd$}&\multirow{2}{*}{R}&&\multirow{2}{*}{F}&&GCN+GRU\\
    &&&&&&&&&&&&+AM\\
    \hline
    MUCas \cite{ijis/ChenZZB22}&&&&&&&$\surd$&R&&F&&GCN+AM\\
    \hline
    CasGAT \cite{iconip/WangL20a}&&&$\surd$&&&&$\surd$&R&&F&&GAT+LSTM\\
    \hline
    \hline
    Cascade2vec \cite{HuangWZ19}&&&$\surd$&&&$\surd$&$\surd$&R&&F&&GNN+Bi-GRU\\
    \hline
    CoupledGNN \cite{CaoSGWC20}&$\surd$&&&&$\surd$&&&R&&C&&GNN+AM\\
    \hline
    Cascade2vec \cite{HuangWZ19}&&&$\surd$&&&$\surd$&$\surd$&R&&F&&GNN+Bi-GRU\\
    \hline
    CoupledGNN \cite{CaoSGWC20}&$\surd$&&&&$\surd$&&&R&&C&&GNN+AM\\
    \hline
    infGNN \cite{WuH020}&$\surd$&&$\surd$&&$\surd$&&&R&F&&&GNN+GM\\
    \hline
    \cite{ShangZWLCSL21}&&&$\surd$&&&&$\surd$&R&F&&&GNN+LSTM\\
    \bottomrule
    \multicolumn{13}{>{\footnotesize\itshape}r}{Continue on the next page}
\end{tabular}
}
\end{table}
}

{\renewcommand{\arraystretch}{1.37}
    \begin{table}[hptb]
\ContinuedFloat
    \centering
\caption{Summary of different methods (cont.)}
\begin{tabular}{c||M{1.8em}|M{1.8em}|M{1.8em}|M{1.8em}|M{1.8em}|M{1.8em}|M{1.8em}||M{2.1em}||M{1.8em}|M{1.8em}|M{1.8em}||c}
    \toprule
     \multirow{2}{*}{Method} & \multicolumn{7}{c||}{Inputs} &\multirow{2}{*}{C/R} &\multicolumn{3}{c||}{Level}  & \multirow{2}{*}{GRL Model}\\
    \cline{2-8}
    \cline{10-12}
     & C & NC & T & CT & SN& CS & CG & & Mi &Ma &Me & \\
    \hline
    \hline
    RMTPP \cite{DuDTUGS16}&$\surd$&&$\surd$&&&&&R&F&&&RNN\\
    \hline
    ERPP \cite{XiaoYYZC17}&$\surd$&&$\surd$&&&&&R&F&&&RNN\\
     \hline
     CYAN-RNN \cite{WangSLGC17}&$\surd$&&$\surd$&&&&&R&F&&&RNN+AM\\
     \hline
    SNIDSA \cite{WangCL18}&$\surd$&&&&$\surd$&&&R&C&&&RNN+AM\\
    \hline
    recCTIC \cite{Lamprier19}&$\surd$&&$\surd$&&&&&R&F&&&RNN\\
    \hline
    CS-RNN \cite{LiuWS19Commu}&$\surd$&&$\surd$&&&&&R&F&&$\surd$&RNN\\
    \hline
    DeepCas \cite{LiMGM17}&&&&&$\surd$&&$\surd$&R&&C&&GRU+AM \\
     \hline
    DCGT \cite{LiGM18}&&&&&$\surd$&&$\surd$&R&&C&&GRU+AM \\
    \hline
    DeepHawkes \cite{CaoSCOC17}&&&&$\surd$&&&&R&&C&&GRU+PM\\
    \hline
    NT-GP \cite{LiuWL19}&$\surd$&&$\surd$&&&&&R&&F&&GRU\\
    \hline
    VaCas \cite{infocom/0002XZTZ20}&&&$\surd$&&&&$\surd$&R&&F&&Bi-GRU\\
    \hline
    CasFlow \cite{xu2021casflow}&&&&&$\surd$&&$\surd$&R&&F&&Bi-GRU\\
   \hline
   \multirow{2}{*}{TempCas \cite{aaai/TangLHXZS21}}&&&&&&\multirow{2}{*}{$\surd$}&\multirow{2}{*}{$\surd$}&\multirow{2}{*}{R}&&\multirow{2}{*}{F}&&Bi-GRU+LSTM\\
   &&&&&&&&&&&&+CNN+AM\\
   \hline
   \cite{LiuBZTX20}&&&$\surd$&&&&$\surd$&R&&C&&Bi-GRU+AM\\
    \hline
    DLIC \cite{chen2021information}&$\surd$&&$\surd$&&&&&R&F&&&GRU+GAT\\
    \hline
    Topo-LSTM \cite{WangZLC17}&$\surd$&&$\surd$&&$\surd$&&&R&C&&&LSTM\\
    \hline
    DeepDiffuse \cite{IslamMAPR18}&$\surd$&&$\surd$&&&&&R&F&&&LSTM+AM\\
    \hline
    LSTMIC \cite{gou2018learning}&&&$\surd$&&&$\surd$&&C&&&&LSTM+PM\\
    \hline
    PATH \cite{manco2018predicting}&$\surd$&&$\surd$&&&&&R&F&&&LSTM\\
    \hline
    DMT-LIC \cite{ChenZ0TZZ19}&$\surd$&&&&$\surd$&&&R&C&C&&Bi-LSTM+GAT\\
    \hline
    DLAM \cite{Yuan18arxiv}&&&$\surd$&&&$\surd$&&R&&F&&LSTM+AM\\
    \hline
    CPMHSA \cite{Liu0J0S20}&$\surd$&&&&$\surd$&&&R&C&&&Transformer\\
    \hline
    GSAN&&&\multirow{2}{*}{$\surd$}&&&&\multirow{2}{*}{$\surd$}&\multirow{2}{*}{R}&&\multirow{2}{*}{F}&&Transformer\\
    \cite{www/Huang00ZZ20}&&&&&&&&&&&&+AM+PM\\
    \hline
    \hline
    FOREST \cite{YangTSC019}&$\surd$&&&&$\surd$&&&R&C&C&&GRU+RL\\
    \hline
    E-FOREST \cite{yang2021full}&$\surd$&&$\surd$&&$\surd$&&&R&F&F&&GRU+RL\\
    \bottomrule
    \end{tabular}
\end{table}
}

Table \ref{tab:freq} lists the details of GRL models that we review in this section. Basically, its information can be divided into four parts: input, task formulation, prediction level, and representation model. We include seven types of input data here: C, NC, T, CT, SN, CS and CG, where C and NC represent the cascade with and without activation ordering, respectively. T represents the activation time series of the cascade. CT and SN are short for cascade diffusion trajectory and the underlying social network, respectively. CS and CG represent cascade size and cascade graph, respectively. Task formulation includes two options: classif\mbox{i}cation (C) and regression (R). For the prediction level, it contains three types: micro-level (Mi), meso-level (Me) and macro-level (Ma), and we further categorize micro-level and macro-level into coarse-grained (C) and f\mbox{i}ne-grained (F). For GRL models, we try to include all techniques that are used in a paper along with the main model, which may include attention mechanism (AM) and pooling mechanism (PM) technique.

\begin{figure}[htbp]
  \centering
  \includegraphics[width=0.5\linewidth]{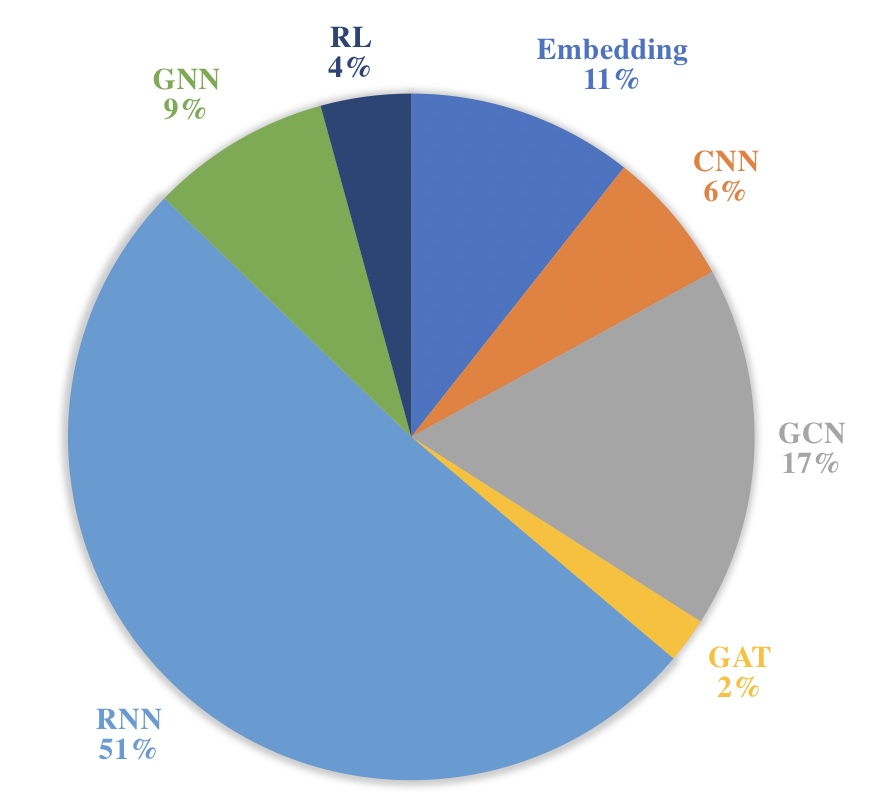}
  \caption{GRL model distributions of reviewed papers.} 
  \label{distribution}
\end{figure}

From Table \ref{tab:freq}, we observe that most of these reviewed papers use activation time as input since it provides temporal features of cascade diffusion for the model. Accordingly, many works apply RNN-based models as the main model to capture these important temporal features and to predict the next activated user and its activation time. This can be seen from the model distribution from Fig. \ref{distribution}.  Another observation is most of these reviewed papers study regression task and only a few of them consider the classf\mbox{i}cation task. Lastly, since different pairs of nodes usually have distinct influence weights, many of these models utilize attention mechanism to capture this signif\mbox{i}cant property between different pairs of nodes.

\section{Embedding-based Methods}\label{Embedding}
In this section, we will talk about the embedding-based methods for popularity prediction problem. Embedding-based or sequence-based methods incorporate the graph information by random walks and maximize the probability of observing neighborhood (context) of a node given its embedding.

\subsection{Random walks}

\begin{definition}[Random walk on graph]
Random walk on graph is a sequence of nodes sampled randomly from the graph. It starts from some node $u$ and travels to one of $u$'s neighbors $v$ with some probability. Then one of $v$'s neighbors is selected and this process is continued. For random walk with restart strategy, there is some probability to return back to the starting node $u$ at each step.
\end{definition}

Two important models based on random walk are DeepWalk \cite{PerozziAS14} and node2vec \cite{GroverL16}. DeepWalk utilizes random walk to sample node sequences and learn  node embeddings by skip-gram method, while node2vec extends the random walk to a biased version by biasing between BFS and DFS. Later, two extensions LINE \cite{TangQWZYM15} and struct2vec \cite{RibeiroSF17} are proposed, which are able to capture more structural information of the graph.

\subsection{Embedding-based models}
Embedding-based approaches for popularity prediction extend IC model, which encode each node into a parameterized vector and then calculate the propagation probabilities between nodes by the learned nodes representations. Bourigault \textit{et al.} \cite{BourigaultLLDG14} embedded nodes in a continuous latent space, and calculated the contamination score of a node activated by the source node through a diffusion kernel function in the latent space. They simplif\mbox{i}ed IC model by assuming active nodes are only activated by the source node. Embedded IC model was proposed in \cite{BourigaultLG16}, which def\mbox{i}ned the propagation probabilities between two nodes by the Euclidean distance between their representations rather than assign a real number. Wang \textit{et al.} \cite{WangSLC15} proposed a model similar to embedded IC model, named LIS model, which can directly learn the user-specif\mbox{i}c influence and susceptibility, and naturally capture context-dependent factors.

Inf2vec \cite{FengCKLLC18} further considered local influence and global user similarity when generating node contexts as an extension. Recently, Panagopoulos \textit{et al.} \cite{panagopoulos2020multi} proposed a model, named IMINFECTOR, which applied the node representations learned from the diffusion cascade to perform model-independent influence spread estimation. For an input node $u$, INFECTOR model utilized a multi-task neural network to simultaneously learn the probability that a node is activated by source node $u$ and the f\mbox{i}nal popularity that $u$ could create.

RNe2Vec \cite{computing/ShangHZPLLX21} estimated the information diffusion probability based on random walk method. Given multiple cascade diffusion trajectories of multiple pieces of information, RNe2Vec f\mbox{i}rst built a repost network and then generated node sequences according to biased random walk. In particular, the probability of choosing next node to visit in random walk is calculated by considering multiple forwarding behaviors: f\mbox{i}rst forwarding, repeated forwarding and self forwarding.

Even though simple embedding-based methods could obtain relatively high quality embeddings to capture structural information of the graph, they are transductive and hard to generalized to unseen data.

\section{Deep Learning Methods}\label{DeepL}
\begin{figure}[htbp]
  \centering
  \includegraphics[width=0.85\linewidth]{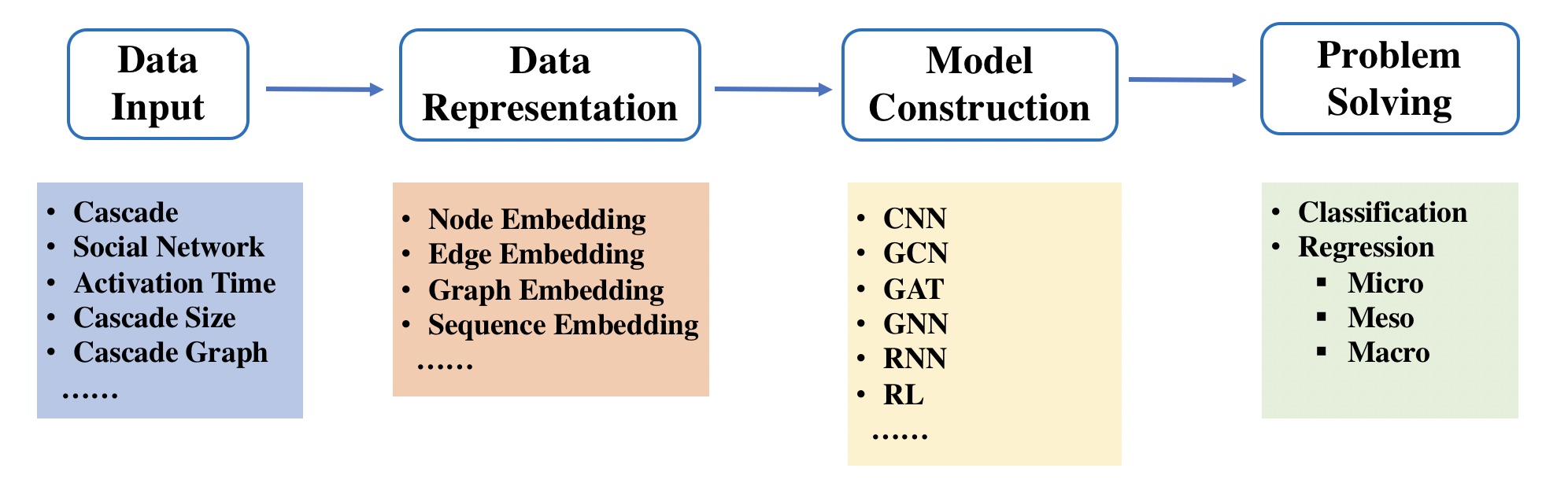}
  \caption{A general framework of using DL models for popularity prediction.} 
  \label{framework}
\end{figure}

DL approaches learn a function, which maps a numeric-form graph to low-dimensional embeddings via optimizing over a large number of expressive neural network functions. In this subsection, we will talk about the literatures related to DL methods. Fig. \ref{framework} shows a general framework of using DL methods to solve popularity prediction. Given some observations of information diffusion, such as cascade and activation time, we usually f\mbox{i}rst extract information from inputs and then embed these information into some low-dimensional latent space, which will be the input to following downstream models. For the selection of DL models, we classify the literatures into six types according to their primary models: CNN, GCN, GAT, GNN, RNN and RL. Finally, the selected DL model will be used to solve the corresponding tasks: classif\mbox{i}cation or regression.  

\subsection{CNN-based models}
CNNs are distinguished from other neural networks by their superior performance in f\mbox{i}nding patterns in images to recognize faces, objects, etc. They are also effective for classifying non-image data, like time series, audio and signal data. Generally, a CNN includes three kinds of layers: convolutional layer, pooling layer and fully connected layer. Convolutional layer is the key building block of a CNN, and majority of computations occur here.  Pooling layers is designed to decrease the number of parameters in inputs, which usually include max pooling and average pooling. Fully connected layer is utilized to do the classif\mbox{i}cation according to the features extracted from previous layers and their different f\mbox{i}lters.

Network-agnostic model Cas2vec \cite{KefatoSBSMG18} integrated CNN with max-pooling technique to capture the information in cascade diffusion time sequence, which can predict whether a cascade will be vital without extracting user features and knowing the underlying network structure. Motivated by two heuristic assumptions, Yang \textit{et al} proposed the NDM model \cite{abs-1812-08933}, which incorporated CNN and AM to solve coarse-grained micro-level prediction. To learn comprehensive node embeddings, HDD \cite{MolaeiZV20} designed a mechanism for extracting features to conduct weighted aggregations of neighbors on different meta-paths. It employed CNN to extract features and LSTM to provide sequential prediction, which solved both the topic diffusion and coarse-grained micro-level prediction.

\subsection{GCN and GAT based models}
GCN performs similar operations as CNN, which learns features from neighborhoods. The key difference between GCN and CNN is that CNN is specially built to operate on Euclidean structured data, while GCN is more general and can handle unordered non-Euclidean data where the number of neighbours of each node is not f\mbox{i}xed and the order of neighbours has no effect on the structure of a node.

End-to-end framework DeepInf \cite{QiuTMDW018} represented both the network structures and diffusion dynamics into a latent space. Given the network and statuses of a node's local neighborhood, DeepInf aimed to predict the activation probability of the node. DeepInf incorporates the graph convolution, network embedding and AM into a unif\mbox{i}ed framework, and its framework has large representative for using GCN or GAT model for popularity prediction problem. 

Chung \textit{et al.} \cite{ChungHC19} designed a model to capture characteristics and tree structure of information diffusion, which simultaneously performed bottom-up and top-down representation aggregation. They utilized a tree pruning strategy on input propagation trees such that more representative nodes are chosen and others are eliminated, to get better tree representation. DyHGCN \cite{YuanLZLZH20} utilized multi-layer GCN to learn the node representation from the repost relations of nodes in timestamped cascade sequence and the social relations of nodes in social network. It applied multi-head attention module, proposed in Transformer model \cite{VaswaniSPUJGKP17}, to model the information of diffusion sequence.

Wang \textit{et al.} \cite{wang2020learning} proposed an end-to-end method, named Multi-Influor, to make the coarse-grained micro-level prediction given the statuses of users in its local subnetwork. Multi-Influor learned multiple representation vectors for each node by incorporating pairwise node interactions, network structures and global similarity comparisons. For the node-level influence representation learning, it mainly used the ramework of DeepInf model \cite{QiuTMDW018}. For the graph-level influence representation learning, it considered three parts to construct the influence context of a node. Except for the local influence context and global influence context used in \cite{FengCKLLC18}, Multi-Influor also considered subnetwork-level influence context. 
 
CasSeqGCN \cite{abs-2110-06836} considered both the structural feature and temporal features. Using the f\mbox{i}xed social network structure and varying node states as input, CasSeqGCN employed GCN to learn the representation of each snapshot of cascade, which accurately modeled the structure information and preserved the temporal order of the spreading. LSTM model is utilized in CasSeqGCN to aggregate node representation and extract temporal information.

Inspired by the DeepInf model \cite{QiuTMDW018}, Leung \textit{et al}. \cite{LeungCMD019} proposed a hybrid algorithm, called HPPNP, which combined two methods for page rank prediction, PPNP and APPNP \cite{abs-1810-05997}, with a hybrid method to enhance the DeepInf model. CasGCN \cite{corr/abs-2009-05152} f\mbox{i}rst proposed a bi-directional convolution technique to represent cascade graph as two graphs: in-coming and out-coming graph, and used GCN to aggregate information not only from in-neighbors of a node but also its out-neighbors. CasGCN learned temporal features by the normalization of diffusion time.                 

MUCas \cite{ijis/ChenZZB22} incorporated GCN, dynamic routing mechanism and AM to learn latent representations of cascade graphs to fully utilize the directional, high-order, positional, and dynamic scales of cascade information. Since the input is only the cascade graph before some specif\mbox{i}c observation time, MUCas designed a novel sampling method to generate sub-cascade graphs based on disjoint time intervals. By modifying the convolution kernel in GCN, MUCas learned node embeddings capturing order, position and direction information, which are then aggregated by dynamic routing.    

The main difference between GCN-based and GAT-based models is that contributions of neighbors to central node in GAT are not pre-determined like in GCN. Instead GAT applies the attention mechanisms to capture the relative weight between two adjacent nodes. Therefore, performance of GAT-based approaches are usually better than GCN-based approaches. CasGAT \cite{iconip/WangL20a} employed GAT to focus on the relationship between nodes and its neighborhood rather than the entire graph. CasGAT utilized LSTM to capture the temporal information in the cascade graph and a non-parametric approach to obtain more proper representations of time decay factors.

\subsection{GNN-based models}
GNN is also useful in dealing with non-Euclidean data and was f\mbox{i}rst proposed in \cite{ScarselliGTHM09}. Later, many variants occurred and one of the most commonly used is inductive framework GraphSAGE \cite{nips/HamiltonYL17}. GraphSAGE leverages node attribute information to eff\mbox{i}ciently generate representations for previously unseen data. The main point of GraphSAGE lies in learning how to aggregate feature information from neighbors. Specif\mbox{i}cally, nodes will f\mbox{i}rst aggregate information from their local in-neighbors. After $K$ iterations, a node will f\mbox{i}nally aggregate the information from its $K$-hop neighbors.

Chen \textit{et al.} \cite{Chen0ZTZZ19} proposed the f\mbox{i}rst GNN-based model for popularity prediction, called CasCN, which leveraged GCN and LSTM to fully capture both the structural and temporal information from the cascade graph with the temporal evolution. CasCN also considered the diffusion direction and time-decay effects. Borrowing the idea from graph isomorphism network and residual networks, Cascade2vec model \cite{HuangWZ19} improved the convolutional kernel in CasCN model and modeled cascades as dynamic graphs to learn cascades representation via graph recurrent neural networks.

Cao \textit{et al}. \cite{CaoSGWC20} designed a novel model, called CoupledGNN, based on the idea that the activation of a node is intrinsically governed by two key factors, i.e., neighbors state and the spread of influence along social networks. CoupledGNN utilized two GNNs coupled through two gating mechanisms to describe the interplay between the spread of influence and node activation states.  Particularly, one GNN, called state GNN, is applied to model diffusion of interpersonal influence, gated by nodes activation states. The other GNN, named influence GNN, captures update of activation state of each node through interpersonal influence from their neighbors, and is used to capture the influence spread on networks. 

Wu \textit{et al.} \cite{WuH020} proposed a model, called infGNN, for diffusion prediction by personalized GNN, which is used to model the interactions between structural dependencies and temporal dynamics. In particular, infGNN generated node representation by considering both the local and global influence contexts and using gated mechanism. Shang \textit{et al.} \cite{ShangZWLCSL21} proposed a novel framework, which incorporates the GraphSAGE and LSTM methods. To concentrate on important nodes and exclude noises caused by other less relevant nodes, they learned importance coeff\mbox{i}cients for nodes and adopted sampling mechanism in graph pooling process. Temporal features are captured by incorporating the inter-infection duration time information into the proposed model by LSTM.

\subsection{RNN-based models}

So far we have seen the feed forward neural networks to handle the data that are independent with each other, which include input layer, hidden layer and output layer. However, if we have some sequence data or with some temporal information such that one data point is dependent on its previous data points, we need to design the neural network to incorporate the dependencies between these data points. The propose of RNN solves this problem since RNNs are naturally f\mbox{i}t for processing time-series data. Traditional feed forward neural networks assume that there is no dependence between the inputs and outputs, while the output of RNNs is dependent on prior points in the sequence. An RNN has a looping mechanism, and allows the hidden state, a representation of previous states, to flow from one step to the next step. 

Although RNN is able to deal with sequential data, it can not learn long-range dependencies among time steps. Therefore, RNN suffers from the short term memory. To solve this issue, two specialized RNNs are proposed: LSTM and GRU. The key components of LSTM are the cell state and its various gates. Cell state is the "memory" of the network and is expected to reduce the short-term memory effects by carrying related information throughout the sequence processing. GRU gets rid of the cell state and utilizes the hidden state to transform information. Recently, Vaswani \textit{et al.} \cite{VaswaniSPUJGKP17} proposed a model called Transformer, which is proposed and optimized for handling sequential data, and its speed is highly accelerated using multi-head attention mechanism.

\subsubsection{RNN-based models}
 RMTPP \cite{DuDTUGS16} and ERPP model \cite{XiaoYYZC17} applied RNN to capture the nonlinear dependencies on previous infections, which can output the probability distribution of the next activated node and its activation time. Sequential models, such as RNN, are effective for cascade modeling, which do not require prior knowledge of the underlying diffusion model and are flexible to capture complicated dependencies in cascades. Existing sequential models often apply a chain structure to model the memory effect. However, cascade is a diffusion tree and there may be cross-dependence existing in cascade. To capture this property, Wang \textit{et al.} \cite{WangSLGC17} proposed an attention-based RNN model, called CYAN-RNN. They further introduced a coverage strategy to adjust allocation of attention and combat the attention misallocation caused by the memoryless of traditional attention mechanisms, which allows alignments to better reflect the propagation structure.
 
SNIDSA \cite{WangCL18} adopted RNN to capture the historical sequential diffusion, and utilized AM to model the structure dependency among nodes. A specially designed gating mechanism is proposed to incorporate the structural and sequential information. recCTIC model \cite{Lamprier19} was the f\mbox{i}rst Bayesian topological RNN model for sequences with tree dependencies, which is built upon continuous-time IC model and are able to extract full diffusion paths from sequential activation observations via black-box inference.

CS-RNN model \cite{LiuWS19Commu}  was proposed to make the f\mbox{i}ne-grained micro-level prediction and to predict the probability that the next activated node belongs to a specif\mbox{i}c community. The intuition behind the model is the activation of the next node is dependent on both historical sequential statuses of activated nodes and communities through which information spreads. To capture the community influence in cascade, CS-RNN designed a community structure label prediction module to predict community label of the next activated node.
 
\subsubsection{GRU-based models}

DeepCas \cite{LiMGM17} is the f\mbox{i}rst end-to-end DL approach for popularity prediction problem, which takes cascade graph as input and is able to directly learn cascade representations without any feature design. Inspired by DeepWalk \cite{PerozziAS14}, DeepCas f\mbox{i}rst sampled several node sequences based on cascade graph by random walk. Then sampled sequences are fed into a bidirectional GRU (Bi-GRU), where a specif\mbox{i}cally designed AM is used to assemble the graph representation. Subsequently, Li \textit{et al.} \cite{LiGM18} proposed the DCGT model, which extended the DeepCas model by jointly considering the structure and text of cascades. The major difference between DCGT and DeepCas is that DCGT applies a gating mechanism to dynamically fuse the structural and textural representation of nodes. Experiments demonstrate DCGT  signif\mbox{i}cantly improves DeepCas's performance. 

DeepHawkes \cite{CaoSCOC17} leveraged end-to-end DL to capture interpretable factors of Hawkes process such that DeepHawkes can simultaneously possess the high predictive power of DL approaches and high interpretability of Hawkes process. Particularly, the three vital interpretable concepts of Hawkes process: user influence, self-exciting mechanism and time decay effect in information diffusion, can exactly be captured by the three components of DeepHawkes model. Given a cascade diffusion trajectory, DeepHawkes utilized GRU, non-parametric time kernel and sum pooling to aggregate contributions of early adopters. Experiments demonstrate that DeepHawkes outperforms DeepCas model for popularity prediction.

Liu \textit{et al.} \cite{LiuWL19} proposed the NT-GP model for non-topological event propagation, which only takes the timestamped cascade as input and can estimate the cascade size within a specif\mbox{i}c time interval. It learned the topological structure of social network by the time decay sampling method which is incorporated with Bi-GRU to predict the cascade size.
 
VaCas \cite{infocom/0002XZTZ20} framework incorporated graph signal processing, Bi-GRU and hierarchical variational autoencoder techniques to capture information from cascade graph. Particularly, graph signal processing techniques are utilized to generate nodes’ structural embeddings from spectral graph wavelets. VaCas leveraged Bi-GRU to model temporal dependencies of information diffusion and utilized hierarchical variational autoencoder to model uncertainties in information diffusion and cascade size growth. VaCas can capture node-level and cascade-level diffusion uncertainties, as well as the contextualized user behaviors. Later, Xu \textit{et al.}  \cite{xu2021casflow} proposed an extension of VaCas, called CasFlow, which leveraged a hierarchical variational information diffusion model to capture both node-level and cascade-level uncertainties, and learned the posterior cascade distribution by variational inference and normalizing flows.

Tang \textit{et al.} \cite{aaai/TangLHXZS21} proposed a novel framework, named TempCas, which incorporated Bi-GRU, LSTM, CNN and AM to fully exploit features in cascade graph in terms of graph structure and graph size time-series. In particular,  TempCas proposed a heuristic method instead of using random walk to sample full critical paths on cascade graphs, and then utilized Bi-GRU with an attention pooling to obtain cascade paths embedding. TempCas further developed an attention CNN mechanism to capture short-term variation over time on cascade graph size and merge the local features within a f\mbox{i}xed window. Then TempCas utilized LSTM over the attention CNN to learn the historical trend.

Lit \textit{et al.} \cite{LiuBZTX20} proposed an end-to-end framework, which incorporates Bi-GRU and AM to make the coarse-grained macro-level prediction. An attention mechanism, involving intra-attention and inter-gate modules, was designed to eff\mbox{i}ciently capture and fuse the structural and temporal information from the observed period of the cascade. DLIC model \cite{chen2021information} incorporated GAT and GRU models,  and constructed a seq2seq framework for learning spatial-temporal cascade features. Particularly, for user spatial feature, based on social network structure, it learns potential relationship among users by GAT. For temporal feature, DLIC constructed a RNN for learning structural contexts in multiple different time intervals based on timestamp with a time-decay attention.

\subsubsection{LSTM-based models}

To better capture the dynamic structure of cascade, Wang \textit{et al.} \cite{WangZLC17} introduced a new data model, called diffusion topology, and explicitly modeled the dynamic directed acyclic graph structure of cascade by the proposed Topo-LSTM model. Topo-LSTM extends classical LSTM model by structuring the hidden states as a dynamic directed acyclic graph. Islam \textit{et al.} \cite{IslamMAPR18} proposed the DeepDiffuse model by incorporating the AM technique into LSTM model, which can predict not only the next activated node but also its activation time. Experiemental results show that  DeepDiffuse outperforms DeepCas \cite{LiMGM17}, RMTPP \cite{DuDTUGS16} and  TopoLSTM \cite{WangZLC17} methods in terms of predicting the next activated node.

LSTMIC \cite{gou2018learning} utilized LSTM to capture the long-term dependency among time moments of retweeting and directly learned sequential patterns from cascades to predict whether a cascade will outbreak. PATH \cite{manco2018predicting} is proposed to predict the probability that a node will be activated before some timestamp, which employed the LSTM model to learn influence and susceptibility embedding for each node. 

The deep multi-task learning framework DMT-LIC \cite{ChenZ0TZZ19} was proposed for both the coarse-grained microscopic and macroscopic popularity prediction. A novel shared-representation layer is designed in DMT-LIC based on attention and gated mechanism, which can capture both node sequences in the cascade and structure of the underlying cascade graph. One disadvantage of DMT-LIC is that the two scales' prediction are conditionally independent and  macro-level prediction can not make full use of micro-level information.

Yuan \textit{et al}. \cite{Yuan18arxiv} proposed a model, called DLAM,  which employed AM to model the process through which individual items gain their popularity. They analyzed the interpretability of the model with the four key phenomena conf\mbox{i}rmed independently in the previous works of long-term popularity dynamics quantif\mbox{i}cation: the intrinsic quality, the aging effect, the recency effect and the Matthew effect. They also analyzed the effectiveness of introducing attention model in popularity dynamics prediction.

\subsubsection{Transformer-based models}
CPMHSA model \cite{Liu0J0S20} was proposed for the coarse-grained micro-level prediction by the multi-head self-attention mechanism, which is actually a direct application of the Transformer model \cite{VaswaniSPUJGKP17}. GSAN model \cite{www/Huang00ZZ20} captured bi-directional and long dependencies between sub-graphs of cascade graph by incorporating a sequence transformer block and graph transformer block. Specif\mbox{i}cally, to capture local and global features, both node neighbors' attention and graph-level attention are utilized in the block. Then a bi-directional multi-head attention is used to learn the long-dependencies between sub-graphs of cascade graph.

\subsection{RL-based models}
RL is concerned with how agents take actions with respect to environment to maximize the cumulative rewards obtained. Although RL has shown its effectiveness in many artif\mbox{i}cial intelligence tasks and IM problem \cite{manchanda2020gcomb, chen2021contingency}, there are not too many works using RL to solve popularity prediction problem in current literatures. Yang \textit{et al.} \cite{YangTSC019} presented a multi-scale diffusion prediction model, called FOREST, to simultaneously make f\mbox{i}ne-grained micro-level and coarse-grained macro-level prediction. Particularly, RL is applied in FOREST to solve the non-differentiable problem, which can make use of the cascade size information of macro-level in the RNN-based model for microscopic diffusion. \cite{yang2021full} generalized the FOREST model in \cite{YangTSC019}, which can solve the f\mbox{i}ne-grained version of both microscopic and macroscopic popularity prediction problem, and improve the performance of FOREST model.

\section{Discussion}\label{Discussion}
In this section, we will discuss the strengths and limitations of these different models, and outline the challenges and future works of popularity prediction problem.

\subsection{Strengths and limitations}
Instead of being dependent on hand-crafted features like feature-based methods, DL methods do not require heavy feature engineering and can automatically learn the features. Compared with diffusion model-based and generative process-based methods, DL methods do not make any assumptions about the diffusion protocols and can capture the underlying diffusion dynamics and trend of popularity. DL methods are distinguished by their powerful learning ability and relatively simple architecture. Furthermore, DL methods are inductive frameworks and can be easily generalized to unseen data.

Generally, different DL techniques have their distinguished advantages and can capture different aspects of features of information. For example, GraphSAGE aggregates features from $K$-hop in-neighbors after $K$ iterations and the aggregation diffusion from in-neighbors to the center node is kind of similar as the information spread on networks. RNN and its variants have superior performance in capturing the temporal features in the diffusion cascade. More and more works choose to incorporate multiple different DL techniques to capture more features of the cascade. With the achievement of DL techniques in other f\mbox{i}elds, more and more well-designed model will be proposed for popularity prediction problem. 

However, there are still some limitations of DL models. One main disadvantage is that DL models are generally considered as "black box", and theoretical analysis and explanation of these methods are still lacking. Recently, lots of researches are focused on explaining the DL models. Understanding and explaining DL models is very helpful for us to analyze the  diffusion popularity prediction on social networks. Even though there are many works that have been focused on attention mechanism and tried to use attention mechanism to capture the weights or propagation probability on edges, no effective results in explaining the relationship between the attention mechanism and propagation probability on edges.  Another limitation is the computation cost of DL methods. DL methods have much higher computation costs than generative process-based and feature-based methods. The selection of hyper-parameter, tuning procedure, overf\mbox{i}tting risk and other factors sometimes make it diff\mbox{i}cult for engineers to achieve satisfying results.

\subsection{Challenges and future opportunities}
In this subsection, we will discuss some challenges in current works and possible future chances.  Firstly, though many works have been focused on popularity prediction problem, there are still some fundamental questions that are not yet addressed. For example, can we really predict the diffusion and popularity of a cascade? If the answer is yes, then to what extent can a cascade be predicted? How to evaluate the performance of different methods? Currently, it is hard to compare the performance of different models since there is no standardized comparison protocols and general cascade datasets.    

Secondly, there are usually lots of information spreading on a social network, such as advertisements of complementary or competitive products on marketing. One direction can be learning the popularity prediction for one piece of information among multiple competitive information spreading on networks. For example, predicting popularity for misinformation or positive cascade information is paramount in misinformation containment problem.

Last, popularity prediction problem is actually the foundation for influence maximization learning (IML) problem, which aims to solve the IM problem based on the approximated influence spread function learned from some previous diffusion results. However, IML problem is more than solving the popularity prediction problem. However, once popularity prediction problem is solved eff\mbox{i}ciently, we can use these techniques to help us on IML problem and also its variants, such as prof\mbox{i}t maximization problem, topic-aware IML problem, contingency-aware IML problem and so on. There are not so many works focused on using the DL methods to solve the variants of IML problem, which might be another direction for future work.

\section{Conclusion}\label{conclusion}
In this paper, we provide a comprehensive review of recent literatures using graph representation learning methods for information diffusion modeling and popularity prediction problem. Firstly, we introduce some basic def\mbox{i}nitions and several commonly used methods for popularity prediction problem. Then we classify the existing works for popularity prediction problem according to their primary methods and discussed the advantages and disadvantages of these methods. The challenges of using graph representation learning based methods to solve the popularity prediction problem and some possible future directions are also discussed.

Except for the primary models, there may be some other classif\mbox{i}cation standards to categorize these literatures. For example, we could classify the literatures according to the features that the model uses. The features include the local or global structure of the graph, user feature and the temporal feature. However, this classif\mbox{i}cation standard may not be good since many works may use multiple kinds of the features and it's hard to distinguish them clearly. For future work, with the fast development of the deep learning models in other f\mbox{i}elds, we can use the state-of-the-art model to solve the popularity prediction problem. Since predicting the next user that will be activated can also be considered as the inference problem, perhaps we can use the variational inference method to solve it, which might also be one of the direction of future work.

\section*{Acknowledgment}
This work is supported in part by NSF under grants 1907472 and 1822985; and Start-up Fund from BNU at Zhuhai under grant 310432104.


\bibliographystyle{plain} 
\bibliography{reference}

\nocite{xia2021graph}

\end{document}